\documentclass[runningheads]{llncs}
\usepackage[T1]{fontenc}
\usepackage{graphicx}
\usepackage{color}
\usepackage{xcolor}
\usepackage{algorithm}
\usepackage{algpseudocode}
\usepackage{multirow}
\usepackage{url}
\usepackage{booktabs}

\newcommand\red[1]{\textcolor{black}{#1}}
\newcommand\brown[1]{\textcolor{black}{#1}}

\begin{document}
\title{Decoding Demographic \textit{un-}fairness\\ from Indian Names}
\titlerunning{Decoding Demographic \textit{un-}fairness from Indian Names}
\author{ Medidoddi Vahini \and
Jalend Bantupalli \and
Souvic Chakraborty \and
Animesh Mukherjee}
\authorrunning{Vahini et al.}

\institute{Indian Institute of Technology, Kharagpur, West Bengal, India \\
\email{\{vahinimirididdi, jalend.bantupalli, chakra.souvic, animeshm\}@gmail.com}}
%
%
\maketitle              
%
\begin{abstract}
  Demographic classification is essential in fairness assessment in recommender systems or in measuring unintended bias in online networks and voting systems. Important fields like education and politics, which often lay a foundation for the future of equality in society, need scrutiny to design policies that can better foster equality in resource distribution constrained by the unbalanced demographic distribution of people in the country. 

We collect three publicly available datasets to train state-of-the-art classifiers in the domain of gender and caste classification. We train the models in the Indian context, where the same name can have different styling conventions (Jolly Abraham/Kumar Abhishikta in one state may be written as Abraham Jolly/Abishikta Kumar in the other). Finally, we also perform cross-testing (training and testing on different datasets) to understand the efficacy of the above models.

We also perform an error analysis of the prediction models. Finally, we attempt to assess the bias in the existing Indian system as case studies and find some intriguing patterns manifesting in the complex demographic layout of the sub-continent across the dimensions of gender and caste.

\end{abstract}
%
%

\section{Introduction}

The \textit{name} of a person can convey various demographic features of the individual. This demographic information plays a crucial role in multiple studies related to racial inequality, recommendation systems, biomedical studies, hate-speech target identification, group sentiment analysis etc.~\cite{racepredict,Ambekar2009NameethnicityCF,s1,s2,s3}. Consequently, much work has been done on demographic classification and a variety of online web APIs and tools capable of predicting demographics from user name~\cite{web1,web2,web3}) exist.
Most of this research work, however, is focused on US demographics \cite{racepredict,racebert}, and many of the works \cite{web1,web2,web3,racepredict} build classifiers that require a proper division of the first name, middle name, and last name to work. So, we introduce caste and gender-specific datasets on Indian demographics, which hosts one-seventh of the world population. Indian names vary significantly over states when compared to other countries due to high religious, ethno-demographic and linguistic variance\footnote{\url{https://www.britannica.com/place/India/Indo-European-languages}}. Also, Indian names do not always fall under the division of first/middle/last name, primarily because the name of a person may contain the name of their ancestors (e.g., Avul Pakir Jainulabdeen Abdul Kalam, former President of India - here the first name of the person is Abdul). Indian names may also contain the name of father in the name of a woman. This can make gender or race detection difficult in faulty name segmentation.

So, to achieve a realistic outcome for name to demography detection, we train our classifiers as an end-to-end sequence classification task, overcoming the need for a segmentation model. The Indian society is divided by gender and caste, unlike the US demographics, divided by races. Thus we focus on caste and gender prediction in this study. To summarize, the objective of this study is to predict the gender and the caste from any complete Indian name overcoming the need to build a segmentation model. We list down our contributions in this paper below.

\begin{enumerate}
    \item Toward fulfilling the classification objective, we build large datasets acquiring already public data of India-wide examination records and parsing electoral rolls containing over 7.63 million unique names.
    \item We demonstrate the efficacy of our model through several case studies and make interesting observations about caste and gender based discrimination in India both online and on the ground. 
    \item We show that there has been an upward trend of participation in competitive exams among women and backward classes over the years. We perform a multi-dimensional study to understand the nuance of caste and gender-based discrimination present in Indian society to help policymakers make data-driven choices.
    \item We perform state-wise chronological analysis to understand the efficacy of discrimination-limiting practices/laws implemented in the Indian states.
    \item We also analyze the Indian social media `Koo' to understand the degree of representation a weaker section of the Indian society has on the web and its improvement over time.
\end{enumerate}

\noindent We have opensourced our codebase encouraging further research\footnote{\url{https://github.com/vahini01/IndianDemographics}}

\vspace{-3mm}
\section{Related work}
\vspace{-3mm}

\noindent\textbf{Gender classification}: Hu et al (2021) \cite{2021_1} inferred gender from user first names (US data) using character based models. They further conclude that using complete names results in better prediction results compared to only first names. \cite{tangetall} (2011) inferred gender from Facebook data with an accuracy of 95.2\% for users in the NYC area using their first names. Muller et al (2016) \cite{2016_1} inferred gender from Twitter usernames with 80.4\% accuracy. Tripathi and Faruqui (2011) \cite{svmiit} presented a SVM based approach for gender classification of Indian names using n-gram suffixes and features based upon morphological analysis, obtaining a F1-score of 94.9\%. We refer the reader to Kruger et al (2019) \cite{2019_1} for a comprehensive survey on gender detection models based on textual data.  There are also a few commercial APIs available for gender detection: Gender API~\cite{web1}, Onograph API~\cite{web3}, Genderize API~\cite{web4}. We have used these APIs as baselines.  

\noindent\textbf{Ethnicity classification and demographic bias}: \brown{Classifying the ethnic category, one of the most telling demographic property of a user, provides an essential data add-on for social science research. Other important applications include biomedical research, population demographic studies, and marketing toward a specific group of individuals \cite{Ambekar2009NameethnicityCF}, \cite{2012_1}. Despite numerous applications, ethnic information of users are often not directly available.} 

To bridge this gap, Sood et al.(2018) \cite{racepredict} made use of registered voters in Florida to infer race and ethnicity from names obtaining a F1-score of 0.83 using LSTMs. Ambedkar et al. \cite{Ambekar2009NameethnicityCF} presented a model that classifies names into 13 cultural/ethnic groups with data extracted from Wikipedia. Giles et al. \cite{2012_1} proposed a name-ethnicity classifier that identified ethnicity from personal names in Wikipedia with 85\% accuracy.

\noindent\textbf{The present work}: \brown{Studies on computational bias in Indian datasets are rare due to unavailability of good published datasets\cite{mediabias}. In this work, we specifically focus on India and attempt to quantify bias in diverse Indian datasets across the two major dimensions Indian Society is divided along- caste and gender. To this purpose we collected \textit{multi-year} data from (i) electoral records of different Indian states, (ii) data corresponding to the India wide $10^{th}$ and $12^{th}$ standard examination, (iii) data corresponding to the top Indian engineering and medical entrance examinations and (iv) data from one of the fastest growing Indian social network. We use pre-trained transformer models to obtain better gender and caste classification performance and perform various case studies using the prediction from the best models to gain insights into the underlying demographic biases based on gender and caste in India.}

\vspace{-5mm}
\section{Datasets}
\vspace{-3mm}
\label{sec:datasets}

We use datasets for two purposes: training models and conducting case studies.
We collected three massive datasets\footnote{Detailed stats are available in Appendix \ref{sec:appendix}} to gather training data on diverse Indian names from the Central Board of Secondary Education (CBSE), the All India Engineering Entrance Examination (AIEEE), and the Electoral Rolls (ER). 

To conduct case studies on the social media data, we resorted to the Koo social network\footnote{\url{https://www.kooapp.com/}}. For educational data in addition to the multi-year CBSE (standard X and XII) and the AIEEE data mentioned earlier we also included the All India Pre Medical Test (AIPMT) data. 

\noindent\textbf{CBSE dataset}:

\noindent\textit{Data for training models} -- \brown{The Central Board of Secondary Education (CBSE) keeps a record of all students' grades\footnote{https://resultsarchives.nic.in}. We scraped a sample of about 100K records from their website for the 2014 and 2015 academic years. CBSE data includes information such as the student's name, father's name, mother's name, and grades. It comprises information from students in the 12$^\textrm{th}$ grade during the previous years. In this dataset, the gender labels for students are not available. However the name of the father and the mother of every student are present. This gives us an easy way to get the names and the corresponding gender labels.} 

\noindent\textit{Data for case study} -- \brown{We collected CBSE grade 10 student data from 2004 to 2010 and CBSE grade 12 data from 2004 to 2012 to conduct the case studies.  The number of unique names in the CBSE grade 10 and grade 12 datasets are 70.09\% and 73.12\%  respectively.}

\noindent\textbf{AIEEE dataset}:

\noindent\textit{Data for training models} -- \brown{The All India Engineering Entrance Examination (AIEEE) is a national examination for admission to engineering colleges. The AIEEE data records\footnote{maintained in the same website as CBSE} used for training corresponded to the years 2009, 2010, and 2011. It includes the students' names, state and caste categories - general/reserved (i.e., OBC/SC/ST)\footnote{\url{https://en.wikipedia.org/wiki/Scheduled_Castes_and_Scheduled_Tribes}}, the fathers' and the mothers' names.}

\noindent\textit{Data for the case study} -- The marksheets of students for AIEEE exams spanning the years 2004 to 2011 are randomly sampled and gathered for the case studies.

\noindent\textbf{ER dataset}: Electoral roll data is gathered from the electoral roll websites of each state government. We collected only English language data from these rolls. We show the states considered for the gender classification in Table~\ref{tab:train_data}(Appendix).

\noindent \textbf{AIPMT dataset}: The All India Pre-Medical Test (AIPMT) is a test for admission to medical schools in India. The AIPMT data obtained spans over the years 2004 to 2011. This dataset is solely used to conduct case studies and provides information on 435,288 students with 327,665 (75.27\%) unique names.

\noindent\textbf{Social media dataset}: Apart from educational data for our case studies, we also gathered data from  Koo\footnote{https://www.kooapp.com/feed} which is a rapidly growing social network in India. For our study we have used the data of all Koo users that has been recently released by \cite{koo}.
 We applied our models to this dataset and analyzed the degree of representation based on caste and gender, as shown in section \ref{sec:casestudies}.
 
\vspace{-3.5mm}
\section{Methodology}
\label{sec:methodology}
\vspace{-3.5mm}

We can determine the gender/caste from a user's name using either the first or full name. In India, extracting the first name from the full name is dependent on the state, religion and local culture; for example, the first name appears as the final word in the name in certain states for certain religions. Hence we used a person's full name as input for both gender and caste classification tasks.

\vspace{-6mm}
\subsection{Classification models}

\vspace{-3mm}
\noindent\textbf{Baselines and models}:

We used the top APIs available as baselines: Gender API \cite{web1}, Genderize API \cite{web3}, and Forebears API \cite{web4}. For non-DL models, we use logistic regression and SVM. We use CharCNN and CharLSTM as neural models trained from scratch. We used BERT, mBERT, IndicBERT, and MuRIL as pretrained neural models. Details are added in Appendix \ref{sec:appendixbaselines}.
\vspace{-3mm}

    
\vspace{6mm}
\begin{table}
    \centering
    \caption{Performance of the models(Accuracy) for gender and caste classification. }
    \begin{tabular}{c|ccc|cc}
    \toprule
         & \multicolumn{3}{c|}{\textbf{Gender classification}}  & \multicolumn{2}{c}{\textbf{Caste classification - AIEEE Data}} \\ 
    \midrule
        Model & \multicolumn{1}{c}{CBSE} & \multicolumn{1}{c}{ER} & \multicolumn{1}{c|}{AIEEE} & Complete Name & Name \& State\\
    \midrule
        LR  & 91.03  & 73.55  & 87.38  & 68.71 & 69.62  \\
        SVM  & 93.82  & 46.85  & 85.31 & 61.73  & 64.82\\
        Char-CNN  & 96.18  & 89.74  & 94.13 & 71.57 & 73.21  \\
        Char-LSTM  & 95.81  & 90.41  & 94.72 & 71.61 & 73.38\\
        BERT  & 96.97  & \textbf{92.56} & \textbf{96.06}& 72.62 & \textbf{74.70} \\
        MuRIL  & \textbf{97.07}  & 92.49  & 95.97& 71.91  & 73.79 \\
        IndicBERT  & 96.32  & 91.52 & 94.59 & 70.66  & 72.86 \\
        mBERT  & 96.80  & 92.50  & 95.84& \textbf{73.05} & 74.61\\
    \bottomrule
    \end{tabular}
    \label{tab:gender_caste_results}
\end{table}

\vspace{-5mm}
\section{Experimental setup}

\noindent\textbf{Gender and Caste labels}: Only binary categories -- male and female -- are used for gender classification task. As for the caste, the categories that one finds are \textit{General} (upper caste people who did not face historical discrimination and benefited from the caste system), \textit{Scheduled Caste} (SC: who were discriminated historically), \textit{Scheduled Tribe} (ST: who were out castes and faced the maximum discrimination), and Other Backward Castes (OBC). For the purpose of this study, we divide castes into broad groups: General and Reserved (SC/ST/OBC for whom the government guarantees reservation to ensure a level-playing field). 

\noindent\textbf{Repetition of names}: Many names in our datasets repeat. Thus it is possible that test points (chosen randomly) can overlap with the training points. In order to avoid this, we run our experiments on unique names only. The label for this instance is the majority label of all the individual instances of the name. For our experiments we use a train-test split of 70:30. More details related to dataset division is included in Appendix \ref{sec:appendix}.

\vspace{-2mm}

\section{Results}
\vspace{-3mm}
\noindent\textbf{Main results}: The main results are noted in Table~\ref{tab:gender_caste_results}. We observe that simple ML models like LR and SVM do not perform well. Character based models show greater improvement over LR, SVM for gender detection showing the benefits of the choice of character sequences for this task. Transformer based models perform best for both gender and caste classification with no clear winner among them. Overall, MuRIL does well in gender classification and mBERT in caste classification. 

\begin{table*}
\scriptsize
\begin{minipage}{0.47\textwidth}
\caption{Comparison with baseline APIs. All results are on a held out set of 500 data points.}
\label{tab:baselineModel}
\begin{tabular}{llccc} 
    \toprule
    
     &  \textbf{Model}   & \textbf{ER} & \textbf{AIEEE}   & \textbf{CBSE}  \\ 
    \midrule
    \multirow{3}{*}{APIS} & Gender~\cite{web1}         & 53.2      & 64.0   & 81.0  \\ 
    & Onograph~\cite{web3}           & \textbf{71.46}     & \textbf{82.00}   & \textbf{92.8} \\ 
    & Genderize~\cite{web4}         & 49.79    & 63.86   & 82.38\\
    \midrule
    \multirow{6}{*}{Models} & MuRIL-CBSE  & 74.85 & 89.00 & 97.00 \\
    & MuRIL-ER  & \textbf{93.81} & 94.20 & 97.40 \\
    & MuRIL-AIEEE  & 77.45 & 95.40 & \textbf{97.60}  \\
    & BERT-CBSE  & 77.05 & 86.20 & 97.20 \\
    & BERT-ER  & \textbf{93.81} & 94.20 & 97.00  \\
    & BERT-AIEEE  & 76.25 & \textbf{97.00} & 97.60  \\
    
    \bottomrule
\end{tabular}
\end{minipage}
\begin{minipage}{0.4\textwidth}
\centering
     \caption{\red{Performance in a cross-dataset setting}}
    \label{tab:transferlearning}
    \begin{tabular}{ccccc} 
    \toprule
        \textbf{Model} & \textbf{Train} & \textbf{Test} & \textbf{Accuracy }& \textbf{ F1-Score}\\
        \midrule
        MuRIL & ER & CBSE & \textbf{97.31} & \textbf{97.28} \\
        & ER & AIEEE  & \textbf{95.40} & \textbf{95.31}\\
       & CBSE & ER & 78.03 & 77.93 \\
        & CBSE & AIEEE & \textbf{90.82} & \textbf{90.72} \\
        & AIEEE & ER & 79.47 &  79.35\\
        & AIEEE & CBSE & \textbf{97.03} & \textbf{97.00}\\
        BERT & ER & CBSE & \textbf{97.31} & \textbf{97.28} \\ 
        & ER & AIEEE  & 94.94 &  94.84\\
        & CBSE & ER & \textbf{78.27} & \textbf{78.15} \\
        & CBSE & AIEEE & 89.64  & 89.50 \\
        & AIEEE & ER & \textbf{79.66} & \textbf{79.53}\\
        & AIEEE & CBSE & 96.99 & 96.96\\
    \bottomrule
    \end{tabular}
\end{minipage}
\end{table*}

\noindent\textbf{Baseline APIs}: We have used only 500 unseen instances due to API request limit per day for each of these baselines. For a set of randomly chosen 500 data points, we observe that the transformer based models (MuRIL and mBERT) by far outperform all the three baselines (see Table~\ref{tab:baselineModel}). Among the baselines, Onograph performs the best.

\noindent\textbf{Cross dataset evaluation}
From Table~\ref{tab:transferlearning} we see that models trained on CBSE and AIEEE datasets (having similar pan-India demographics) perform well on each other's test sets. Further, the models trained on the ER dataset perform reasonably well when tested on CBSE and AIEEE datasets but the reverse setups perform poorly due to lesser representation of north-eastern states in CBSE/AIEEE data. 



\vspace{-3mm}
\section{Decoding unfairness across gender and caste lines}
\label{sec:casestudies}
\vspace{-3mm}

We use social media datasets, longitudinal educational records, and electoral roll datasets to identify and quantify the inter-sectional bias caused by caste and gender prejudice. We also display the results over an 8-year period to better understand how government-sponsored social programs and globalization are influencing Indian society and reducing unfairness in resource distribution.

For all these studies, we have used the MuRIL-based model, which is also shown to be one of the top-performing models.

\noindent To understand the state and evolution of discrimination in the current Indian system, we draft the following set of research questions (\textbf{RQ}):

\begin{figure}[t]
    \centering
    \begin{minipage} {0.3\textwidth}
        \includegraphics[width=\textwidth]{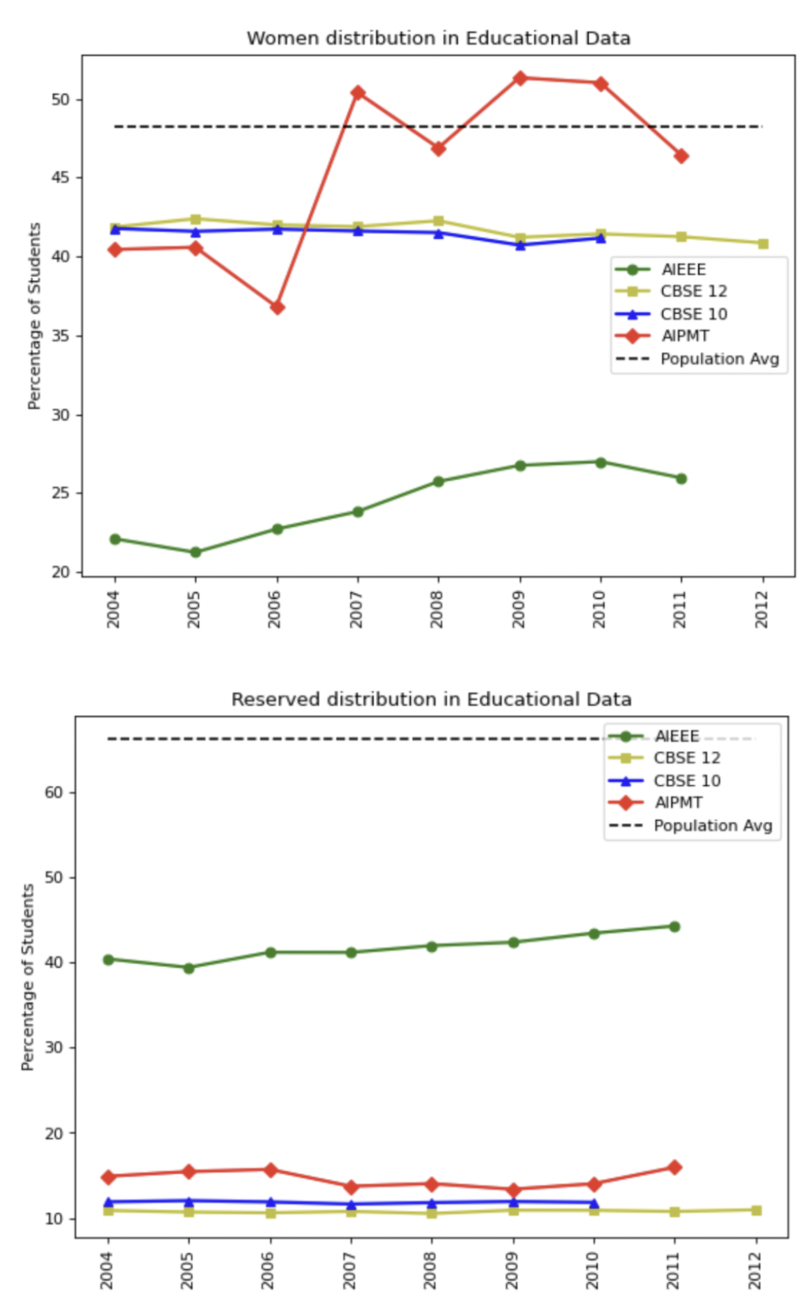}
    \end{minipage}
    \begin{minipage} {0.3\textwidth}
        \includegraphics[width=\textwidth]{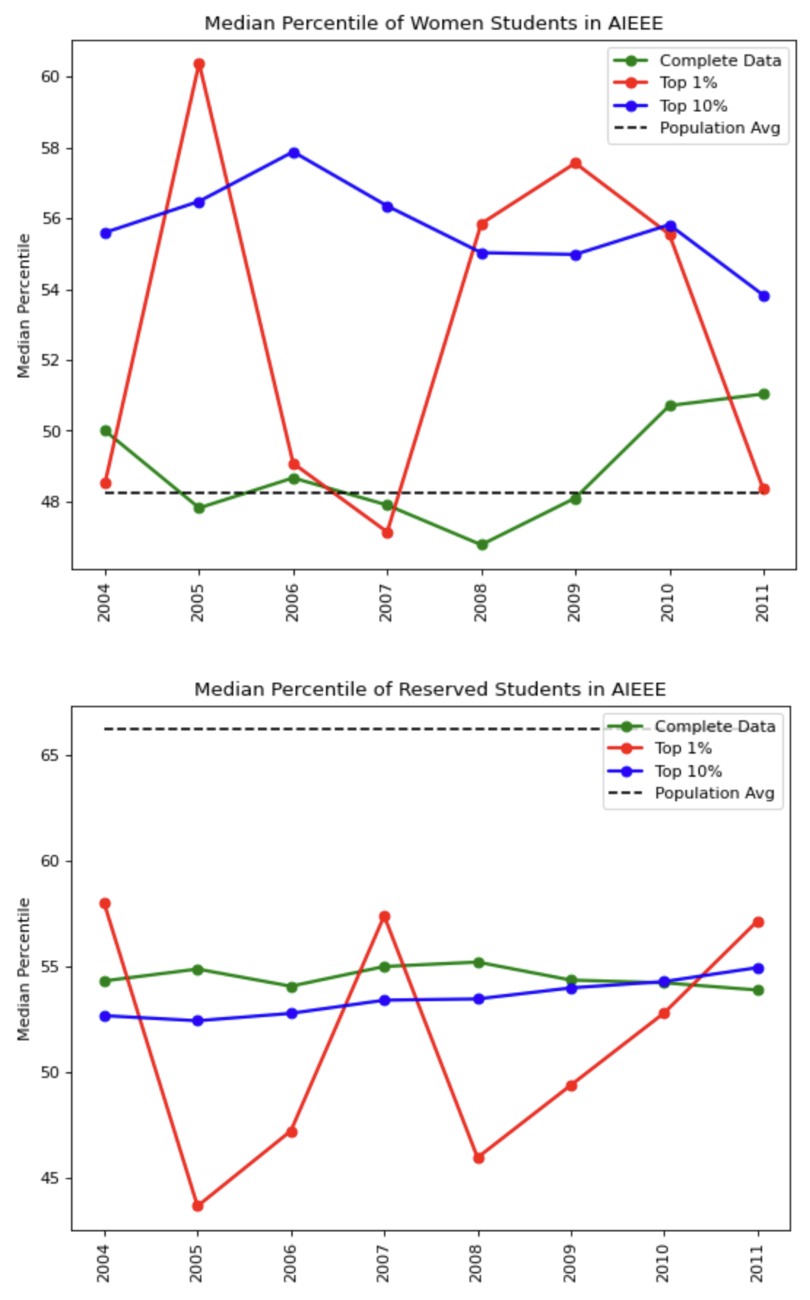}
    \end{minipage}
    \begin{minipage} {0.3\textwidth}
        \includegraphics[width=\textwidth]{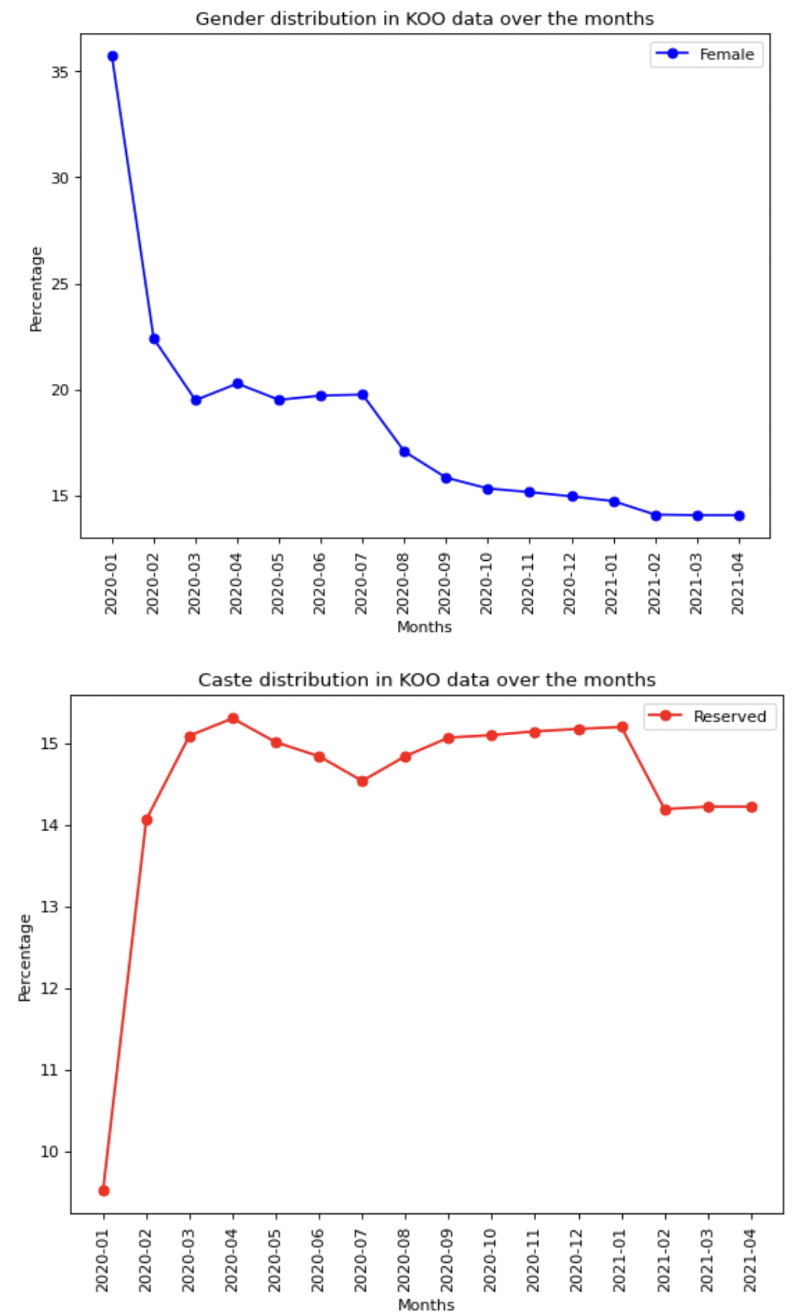}
    \end{minipage}
   \caption{\red{Distribution of women and backward castes are shown in the left most figure. Middle figure displays the median percentile in the same. The right most diagram shows the temporal evolution of women and backward caste people in Koo.}}
   \label{fig:cs}
\end{figure}
\vspace{-2mm}

\begin{itemize}
    \item \textbf{RQ1}: Is the representation of females (no reservation) and backward castes (reserved) in \textit{public education} increasing over time?
    \item \textbf{RQ2}: Is the representation of females (no reservation) and backward castes (reserved) in \textit{competitive engineering/medical entrance exams} increasing over time?

    \item \textbf{RQ3}: Are females and backward caste people less represented in the Indian social networks?
    \item \textbf{RQ4}: How vocal are the females and backward caste voices (\#followers*\#posts) in Indian social media?
\end{itemize}

\subsection{Impact of social bias on education}

\noindent \textbf{RQ1 \& RQ2}: \textit{Is the representation of females (no reservation) and backward castes (reserved) in \textit{public education and competitive exams} increasing over time?}

Indian Government has the promise of equality embedded in its emblem, and the reservation system was introduced as a mechanism to ensure the continuous integration of backward caste people into the mainstream social structure. This reservation system ensures equality of representation in government institutes of higher education and workplaces by reserving some seats for backward class people who have faced historic injustice.

However, a similar system does not exist for women. Hence, we try to understand the system's effect on education for both gender and caste divisions.

In Figure~\ref{fig:cs}, we plot the ratio of women and backward castes in the sampled dataset for each year and each exam. We observe that for AIEEE, women's representation has steadily increased from 2005 to 2010. Also, the engineering exam AIEEE saw much lower participation from women when compared to the medical entrance exam AIPMT or the board exams CBSE 10th and CBSE 12th grade. In the medical examination AIPMT, women's representation has increased and has achieved the population average from 2007. In the CBSE 10th and CBSE 12th grade, we see a stable representation of women over time.

However, the representation of backward castes in the medical entrance exam and both the board exams has remained extremely low all over time. AIEEE, the engineering entrance exam, stands out here with much higher representation from backward castes, and the participation is also slightly increasing with time.

In Figure~\ref{fig:cs} we plot the median percentile attained by women and backward castes for each year and each exam. While the average percentile for the whole dataset will be the 50th percentile, this metric, when applied to a specific community, will tell us the meritocratic position of that community, i.e., whether that community is doing better relative to other communities or lesser and how that position is changing over time concerning specific exams. We have the ranks available only for AIEEE data, so we try to analyze only this data. We see that women have achieved equality in percentile and are doing better than men in later years. However, the same is not valid for backward caste students. The above percentile changes show need for exam-specific policy modifications. For women, the inequality is mostly in participation, and women need more encouragement to participate in competitive exams, whereas backward class students need better support to score higher in the competitive exams.

\vspace{-2mm}

\subsection{Impact of social bias on online media and web representation}
\vspace{-2mm}

\noindent \textbf{RQ3}: \textit{Are women and backward caste people less represented in the Indian social networks?}

We analyze the Indian social network `Koo'\footnote{\url{https://www.kooapp.com/}} to investigate the extent of representation of women and lower caste people in this network. We use the massive dataset with 4 million usernames and metadata from Koo provided by \cite{koo}. Using our classifier, we predicted the caste and the gender of these names. In Figure ~\ref{fig:cs}, we plot the ratio of women and backward castes in the network for each month across Koo's existence. We observe that upper caste people of India mostly occupied Koo during its inception stage; however the situation quickly changed over time with a slightly higher representation from the lower caste people. On the other hand, women's distribution continuously decreased in this Indian social network. 


\noindent \textbf{RQ4}: \textit{How vocal are the women and backward caste voices (\#followers*\#tweets) on Indian social media?}

We quantify the overall `voice' of a group by the sum of the \#followers*\#tweets for each person in that group. As we do not have the number of followers data for each month, we only compute the aggregate statistics at the end of the last month in the dataset. The ratio of male voice to the female voice in the network is 3.59. Similarly the ratio of the general to the backward class voice is 10.47. These results demonstrate a striking inequality. Further we observe that a random message posted on the platform is 37.58 times more likely to be from a forward caste male than a backward-caste female.

\section{Conclusion and future work}
The paper introduced various large-scale datasets of Indian names and extensively explored the possibility of gender and caste detection from these names. We showed that the state-of-the-art APIs do not perform well on the task of gender detection from Indian names; in contrast, the recent transformer based models performed extremely well in this task. Further, to the best of our knowledge, this is the first large-scale caste classification task undertaken to understand the existing demographic disparities in across India. Through a series of rigorous case studies we have shown the gender and caste based biases that exist in basic and higher education as well as in the representation in social media. We have also opensourced our codebase for further research and contribution. In future, we will like to consider more caste varieties and data from all states for a nuanced evaluation.

\bibliographystyle{splncs04}
\bibliography{sample-base}
\appendix
\section{Appendix}
\label{sec:appendix}

\subsection{Dataset statistics}

Table \ref{tab:train_data} displays the dataset stats.

\begin{table}[!th]
\centering
\caption{The table below contains information on datasets that are used to train models and conduct case studies.}
\label{tab:train_data}
\begin{minipage}{0.45\textwidth}

    \begin{tabular}{lll}
    \toprule
        \multicolumn{3}{c}{\textbf{Gender classification}} \\
    \midrule
    \textbf{Data (full)}     & Female & Male \\
    \midrule
        CBSE & 194423 & 194413 \\
        ER & 10405236 & 11632598 \\
        AIEEE & 358522 & 358522\\
    \midrule
    \textbf{CBSE-breakup}    & \multicolumn{1}{l}{Female}  & Male    \\
    \midrule
        2014 &  25779 & 31573 \\
        2015 & 51744 & 63434 \\
    \midrule
    \textbf{ER-breakup} & \multicolumn{1}{l}{Female}  & Male    \\ 
    \midrule
     Daman     & 53391  & 53605  \\ 
     Manipur   & 580415  & 589948 \\ 
     Meghalaya & 748820  & 737951 \\ 
     Nagaland  & 253274 & 295039  \\ 
     Arunachal & 292158  & 292544  \\ 
     Delhi    & 966324  & 1430743  \\ 
     Sikkim       & 76145 & 88209  \\ 
     Goa   & 372029  & 361380  \\ 
     Mizoram     & 134305 & 158144 \\ 
     \midrule
     \textbf{AIEEE-breakup}    & \multicolumn{1}{l}{Female}  & Male    \\
    \midrule
        2009 &  66286 & 84615 \\
        2010 & 70826 & 91687 \\
        2011 & 68965 & 89490 \\
    \bottomrule
    
    \end{tabular}
\end{minipage}
\begin{minipage}{0.45\textwidth}
    \centering
    \begin{tabular}{lll}
    \toprule
        \multicolumn{3}{c}{\textbf{Caste classification}} \\
    \midrule
    \textbf{AIEEE} & Reserved & General \\
    \midrule
        2009 & 47681 & 64892 \\
        2010 & 54703 & 68163 \\
        2011 & 57262 & 65810 \\
    \bottomrule
    \addlinespace
    \toprule
    \multicolumn{3}{c}{\textbf{Case study - education data}} \\
    \midrule
    \textbf{Dataset} & Total & Unique names \\
    \midrule
        AIEEE & 665227 & 525631 \\
        CBSE 10 & 487080 & 341430 \\
        CBSE 12 & 378123 & 276476\\
        AIPMT & 435288 & 327665\\
    \bottomrule
    \addlinespace
    \toprule
    \multicolumn{3}{c}{\textbf{Case study - social media data}} \\
    \midrule
    \textbf{Dataset} & Total & Valid names \\
    \midrule
        Koo &  4061670 & 1761958\\
    \bottomrule
    \end{tabular}
\end{minipage}
\end{table}

\subsection{Baseline APIs and Models}
\label{sec:appendixbaselines}
We used a bunch of APIs available for gender classification as baselines and compared them with the results obtained from our transformer based methods. \\
\noindent\textit{Gender API}~\cite{web1}: Gender-API.com is a simple-to-implement solution that adds gender information to existing records. It receives input via an API and returns the split-up name (first name, last name) and gender to the app or the website. According to the website, it will search for the name in a database belonging to the specific country, and if it is not found, it will perform a global lookup. If it cannot find a name in a global lookup, it performs several normalizations on the name to correct typos and cover all spelling variants.\\
\noindent\textit{Onograph API}~\cite{web3}: OnoGraph is a set of services that predicts a person's characteristics based on their name. It can predict nationality, gender, and location (where they live). The services are based on the world's largest private database of living people, which contains over 4.25 billion people (as of July 2020). According to the documentation, ``OnoGraph's results are the most accurate of any comparable service; and it recognizes around 40 million more names than the nearest comparable service.''\\
\noindent\textit{Genderize API}~\cite{web4}: It is a simple API that predicts a person's gender based on their name. The request will generate a response with the following keys: name, gender, likelihood, and count. The probability denotes the certainty of the gender assigned. The count indicates the number of data rows reviewed to calculate the response.\\

\subsection{Model description}

\noindent\textbf{Logistic regression}: We concatenate the different parts of the name and compute character n-grams. Next we obtain TF-IDF scores from the character n-grams and pass them as features to the logistic regression model. 
   
\noindent\textbf{SVM}: The objective of the support vector machine algorithm is to identify a hyperplane in N-dimensional space (N = the number of features) that categorizes the data points clearly. Then, we accomplish classification by locating the hyper-plane that best distinguishes the two classes. There are several hyperplanes that might be used to split the two groups of data points. Our goal is to discover a plane with the greatest margin or the greatest distance between data points from both classes.

\noindent\textbf{Char CNN}: Character-level CNN (char-CNN) is a well-known text classification algorithm. Each character is encoded with a fixed-length trainable embedding. A 1-D CNN is applied to the matrix created by concatenating the above vectors. In our model, we utilize 256 convolution filters in a single hidden layer of 1D convolution with a kernel size of 7. 

\noindent\textbf{Char LSTM}: A name is a sequence of characters. Like char-CNN, each character of the input name is transformed into trainable embedding vectors and provided as input. Our model employs a single LSTM layer with 64 features and a 20\% dropout layer. 


\noindent \textbf{Transformer models}
\begin{itemize}
    \item We choose BERT for demographic categorization, using full names as inputs because it has proven to be highly efficient in English data sequence modeling.
    \item mBERT is trained using a masked language modeling (MLM) objective on the top 104 languages with the largest Wikipedia.
    \item IndicBERT is a multilingual ALBERT model that has only been trained on 12 major Indian languages\footnote{IndicBERT supports the following 12 languages: Assamese, Bengali, English, Gujarati, Hindi, Kannada, Malayalam, Marathi, Oriya, Punjabi, Tamil, and Telugu.}.  
    IndicBERT has much fewer parameters than other multilingual models.
    \item MuRIL is pre-trained on 17 Indian languages and their transliterated counterparts. It employs a different tokenizer from the BERT model. This model is an appropriate candidate for categorization based on Indian names because it is pre-trained on Indian languages.
\end{itemize}

\noindent\textbf{Hyperparameters}:
\textbf{LR}: learning rate = 0.003, n-gram range = (1-6)\\
\textbf{SVM}: kernel=rbf, n-gram range = (1-6), degree = 3, gamma = scale \\
\textbf{Char CNN}: learning rate = 0.001, hidden layers = 1, filters = 256, kernel size = 7, optimizer = adam\\
\textbf{Char LSTM}: learning rate = 0.001, dropout = 0.2, hidden layers = 1, features = 64, optimizer = adam\\
\textbf{Transformer models}: models = [bert-base-uncased, google/muril-base-cased, ai4bharat/indic-bert, bert-base-multilingual-uncased], epochs = 3, learning rate = 0.00005\\

\subsection{Results}

More detailed results are given in Table \ref{tab:gender_results} and \ref{tab:caste_results}.

\red {\noindent \textbf{Handling of corner cases} : 
As a name can be common across both genders or caste, we use majority voting inorder to label a name with binary label for both gender and caste classification tasks. In case of equality we considered arbitrarily decided labels.} 

\begin{table}
    \centering
    \caption{Performance of the models for gender classification on each dataset. }
    \begin{tabular}{c|cc|cc|cc}
    \toprule
        \multicolumn{7}{c}{\textbf{Gender classification}}  \\ 
    \midrule
        Model & \multicolumn{2}{c}{CBSE} & \multicolumn{2}{c}{ER} & \multicolumn{2}{c}{AIEEE} \\
    \cmidrule{2-7}
         & F1-Score & Accuracy & F1-Score & Accuracy & F1-Score & Accuracy \\
    \midrule
        LR & 90.93 & 91.03 & 73.23 & 73.55 & 87.24 & 87.38 \\
        SVM & 93.69 & 93.82 & 37.91 & 46.85 & 85.12 & 85.31\\
        Char-CNN & 96.12 & 96.18 & 89.72 & 89.74 & 94.54 & 94.13 \\
        Char-LSTM & 95.75 & 95.81 & 90.23 & 90.41 & 94.62 & 94.72 \\
        BERT & 96.94 & 96.97 & \textbf{92.52} & \textbf{92.56} & \textbf{95.99} & \textbf{96.06}\\
        MuRIL & \textbf{97.04} & \textbf{97.07} & 92.45 & 92.49 & 95.90 & 95.97\\
        IndicBERT & 96.28 & 96.32 & 91.48 & 91.52 & 94.48 & 94.59 \\
        mBERT & 96.76 & 96.80 & 92.46 & 92.50 & 95.76 & 95.84\\
    \bottomrule
    \end{tabular}
    \label{tab:gender_results}
\end{table}

\begin{table}
    \centering
    \caption{Performance of the models for caste classification on AIEEE dataset. }
    \begin{tabular}{c|cc|cc}
    \toprule
        \multicolumn{5}{c}{\textbf{Caste classification - AIEEE Data}}  \\ 
    \midrule
        Model & \multicolumn{2}{c}{Complete Name} & \multicolumn{2}{c}{Name \& State}  \\
    \cmidrule{2-5}
         & F1-Score & Accuracy & F1-Score & Accuracy \\
    \midrule
        LR & 68.64 & 68.71 & 69.58 & 69.62  \\
        SVM & 53.82 & 61.73 & 59.58 & 64.82\\
        Char-CNN & 71.18 & 71.57 & 72.74 & 73.21  \\
        BERT & \textbf{71.80} & 72.62 & \textbf{73.99} & \textbf{74.70} \\
        MuRIL & 71.57 & 71.91 & 73.04 & 73.79 \\
        IndicBERT & 69.72 & 70.66 & 72.03 & 72.86 \\
        mBERT & 71.34 & \textbf{73.05} & 73.60 & 74.61\\
    \bottomrule
    \end{tabular}
    \label{tab:caste_results}
\end{table}

\subsection{Error Analysis - Baseline APIs vs Our Models}

Table \ref{tab:erroranalysis} lists some of the best and worst test cases for the best performing baselines and the best performing transformer based models. Both these types of models perform the best when the first name (first word) is a good representative of the gender (e.g., Karishma Chettri). Baselines usually fail in three cases: the presence of parental name or surname (e.g., Avunuri Aruna), longer names where gender is represented by multiple words (e.g., Kollipara Kodahda Rama Murthy), and core Indian names (e.g., Laishram Priyabati, Gongkulung Kamei). The main reason for the better performance of transformer models might be that they are trained on complete names and larger datasets. As a result, they handle the complexity of Indian names. However, both these types of models tend to fail in presence of unusual and highly complicated names (e.g., Raj Blal Rawat, Pullammagari Chinna Maddileti).

\begin{table}
\centering
    \caption{Table listing some common errors by the best performing baselines and the best performing transformer models. Here \textbf{W} stands for wrong and \textbf{C} stands for correct. And \textbf{XX} denotes the model, API results respectively; for e.g., \textbf{WC} lists names where transformer predicted wrong while API predicted correct. The letter in bracket denotes the gender (M for male and F for female). \red{The listed names have multiple instances in the datasets. So none of the names uniquely identify any person} }
    \label{tab:erroranalysis}
    \begin{tabular}{lll} 
    \toprule
        \textbf{Dataset} & \textbf{CC}  & \textbf{CW}  \\ 
        \midrule
        CBSE &  Himanshu Bharatia (M) & Vijay Laxmi Soni (F)  \\
        & Sudha Chaturvedi (F) & Gang Shyam Herau (M)   \\
        ER &  Karishma Chettri (F) & Chingakham Romita (F)  \\
          & Shekhar Sethi (M) & Ramesh Kasarlekar (M) \\
        AIEEE & Suguna (F) & Indra Kumar Singh Bundela (M)  \\
         & Sudeep Agrawal (M) & Avunuri Aruna (F)\\
    \toprule
        \textbf{Dataset} & \textbf{WC}  & \textbf{WW} \\
        \midrule
        CBSE &  Sharmil arora (M) & Raj Blal Rawat (F) \\
        & Ramkanwar gund (F) & Vimal Soni (M)  \\
        ER & Jmod Kyrsian (F) & Embha Lyngdoh (M) \\
        & Esphorlin Thongnibah (M) & Basanta Thapa (F) \\
        AIEEE & Tazeen Husain (F) & Dogin Yapyang (F) \\
        &  Zakki Khan (M) & Pullammagari Chinna Maddileti (M) \\
    \bottomrule
    \end{tabular}
\end{table}

\subsection{\red{Case studies - Values of Median percentile}}

Table \ref{tab:median_percentile_data} displays values that are plotted in the left plot of figure \ref{fig:cs}.
\begin{table}[t]
    \centering
    \begin{tabular}{l|lll|lll}
    \toprule
          Dataset & \multicolumn{3}{c}{Women} & \multicolumn{3}{c}{Reserved}\\
         \midrule
           &  All Data & Top 1\% & Top 10\%  & All Data & Top 1\% & Top 10\% \\
           \midrule
         AIEEE 2004 & 50.00  & 48.53 & 55.60 & 54.31 & 57.81 & 52.67 \\
          AIEEE 2005 & 47.82 & 60.38 &56.48 & 54.87 & 43.68 & 52.43 \\
          AIEEE 2006 & 48.67 & 49.08 & 57.88 & 54.06 & 47.23 & 52.78 \\
          AIEEE 2007 & 47.90 & 47.14 & 56.35 & 54.99 & 57.38 & 53.40 \\
          AIEEE 2008 & 46.78 & 55.85 & 55.03 & 55.20 & 45.97 & 53.46 \\
          AIEEE 2009 & 48.10 & 57.56 & 54.98 & 54.35 & 49.39 & 53.98 \\
          AIEEE 2010 & 50.71 & 55.55 & 55.82 & 54.22 & 52.8 & 54.28 \\
          AIEEE 2011 & 51.04 & 48.35 & 53.83 & 53.88 & 57.14 & 54.94 \\
         \bottomrule
    \end{tabular}
    \caption{Median perctile of Women and Reserved students in AIEEE data}
    \label{tab:median_percentile_data}
\end{table}

\subsection{Case Studies - State wise Results}

To understand state wise distribution of Caste and Gender, we answer following additional research questions(ARQ).

\begin{itemize}
    \item \textbf{ARQ1}: Which states in India have the highest representation of females and backward castes in higher education compared to its population?
    \item \textbf{ARQ2}: Which states in India have been successful in achieving a significant decrease in bias toward females and backward castes over time? Which states are lacking in this aspect?
\end{itemize}

\begin{figure}
    \centering
    \includegraphics[width=\textwidth]{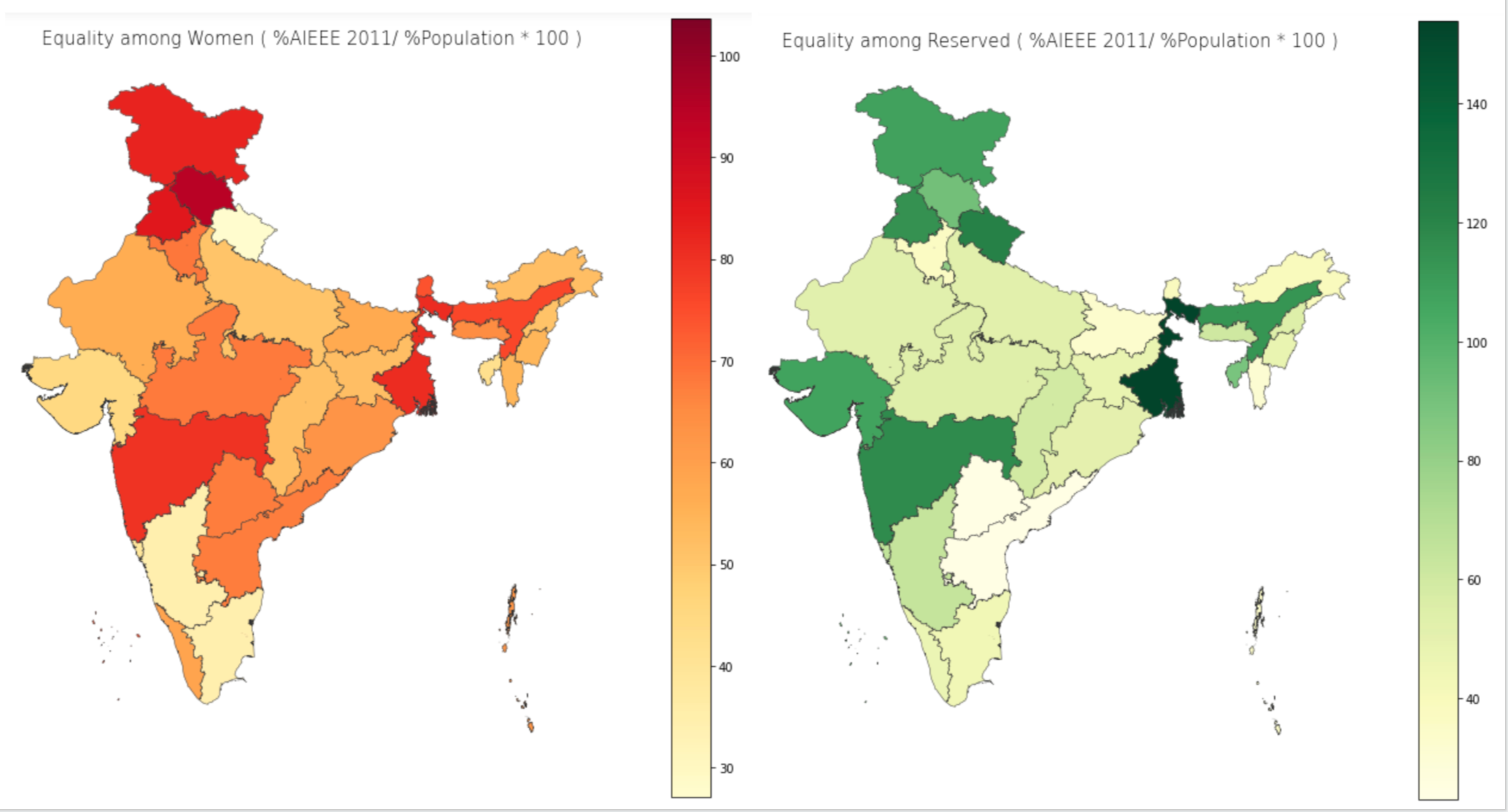} 
    \caption{Population normalized distribution of women and backward caste students in AIEEE 2011 data across Indian states.}
    \label{fig:equalityWomenReservedAIEEE}
\end{figure}
    
\begin{figure}
    \includegraphics[width=\textwidth]{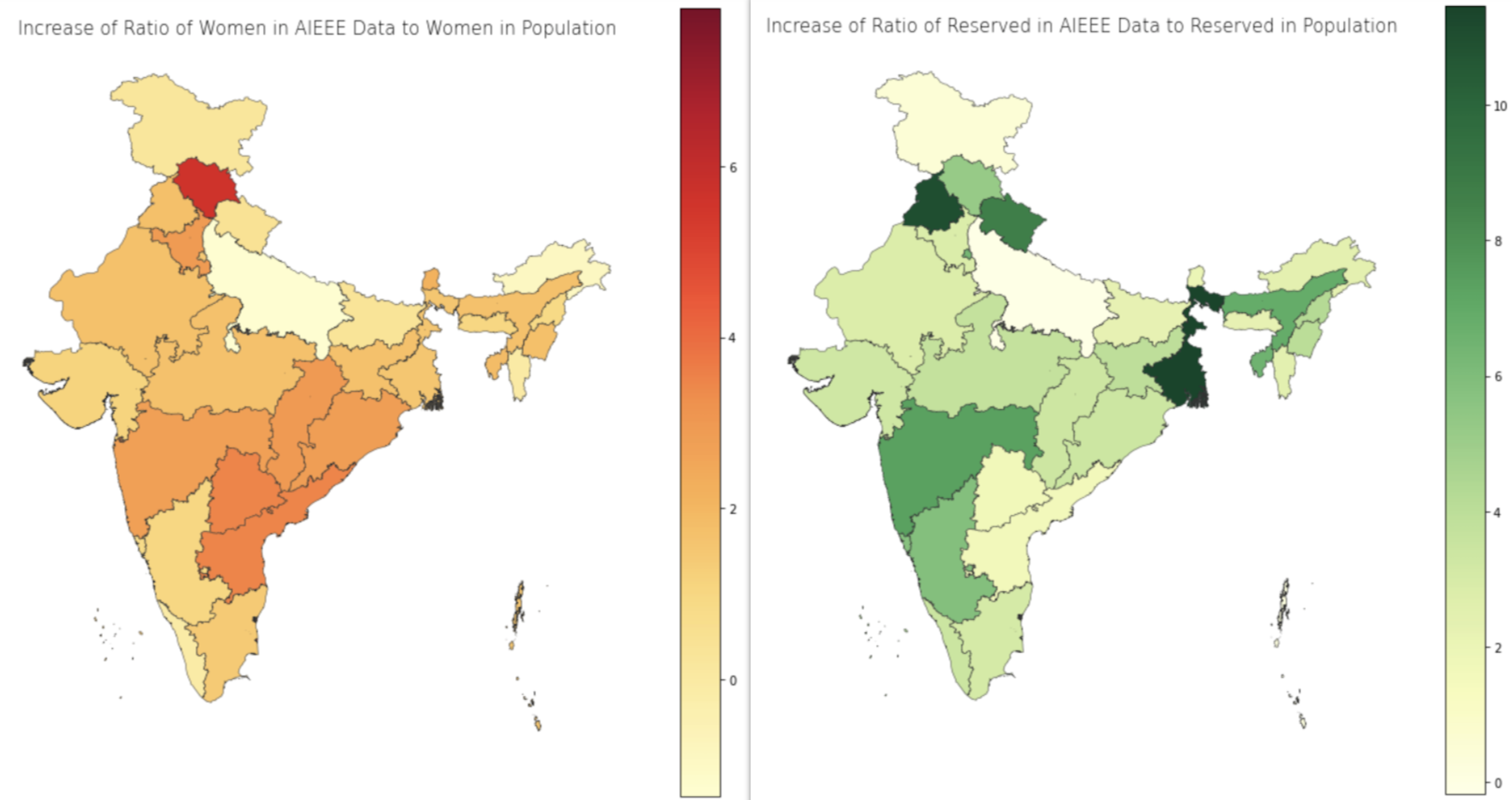} 
    \caption{Rate of change of population normalized percentage of women and backward caste students across Indian states appearing in AIEEE exams.}
    \label{fig:increaseratioWomenReservedAIEEE}
\end{figure}

\noindent \textbf{ARQ1}: \textit{Which states in India have the highest representation of females and backward castes in higher education compared to its population?}

The AIEEE dataset has the state information for each data point. We also collect the state wise population record from Census 2011\footnote{https://en.wikipedia.org/wiki/2011\_Census\_of\_India}. We compute the population normalized fraction of women and backward caste people writing the AIEEE 2011 exam. From the plotted results in Figure~\ref{fig:equalityWomenReservedAIEEE}, we observe that the top states with population normalized higher representation of women writing the AIEEE exam are Jammu \& Kashmir, Himachal Pradesh, Punjab, West Bengal, and Maharashtra. Similarly, the states with population normalized higher representation of backward castes writing the AIEEE exam are West Bengal, Maharashtra, Punjab, Uttarakhand, and Jammu \& Kashmir. We believe that the education policies of these states could act as a suitable guidance to improve the condition of the other Indian states. 

\noindent \textbf{ARQ2}: \textit{Which states in India have been successful in achieving a significant decrease in bias toward females and backward castes over time? Which states are lacking in this aspect?}

One way to measure the reduction (increase) in bias would be to check for the increase (decrease) in the population normalized percentage of women and backward caste over time. To this purpose, we obtained the rate of change of population normalized women and backward class candidates taking the AIEEE exam. For each state, the rate of change is measured as the slope of the best fit line (linear regression) of the year versus population normalized percentage scatter plot.  The year range considered was 2004 to 2011.  


From Figure~\ref{fig:increaseratioWomenReservedAIEEE}, we observe that the most successful states in reducing the gender inequality are Himachal Pradesh, Andhra Pradesh (Seemandhra and Telangana), Haryana and Maharashtra. With respect to reducing caste inequality we find West Bengal, Punjab, Uttarakhand, Maharashtra, Karnataka are the most successful.

\subsection{Distribution of Caste and Gender in Koo}

\begin{table}[!ht]
\centering
\caption{Gender and caste breakup (\%) in the Koo data.}
\label{tab:koo1}
    \begin{tabular}{lll}
    \toprule
    & General & Reserved \\ 
    \cmidrule{2-3}
    Male & 73.26 & 12.66 \\ 
    Female & 12.52 & 1.56 \\ 
    \bottomrule
    \end{tabular}
\end{table}

\begin{table}[!ht]
\centering
\caption{\% users at in the oldest 1\% data sorted by creation date.}
\label{tab:koo3}
    \begin{tabular}{lll}
    \toprule
    & General & Reserved \\ 
    \cmidrule{2-3}
    Male & 67.1 & 12.66 \\ 
    Female & 17.65 & 2.58 \\ 
    \bottomrule
    \end{tabular}

\end{table}

\begin{table}[ht]
    \centering
\caption{\% users at in the most recent 1\% data sorted by creation date.}
\label{tab:koo2}
    \begin{tabular}{lll}
    \toprule
    & General & Reserved \\ 
    \cmidrule{2-3}
    Male & 73.44 & 13.84 \\ 
    Female & 11.22 & 1.50 \\ 
    \bottomrule
    \end{tabular}
\end{table}

\begin{table}[!th]
\centering
\caption{\% users in the top 1\% data sorted by number of followers.}
\label{tab:koo4}
    \begin{tabular}{lll}
    \toprule
    & General & Reserved \\ 
    \cmidrule{2-3}
    Male & 63.87 & 9.00 \\ 
    Female & 24.65 & 2.47 \\ 
    \bottomrule
    \end{tabular}

\end{table}

\begin{table}[!th]
\centering

\caption{\% users in the bottom 1\% data sorted by number of followers.}
\label{tab:koo5}
    \begin{tabular}{lll}
    \toprule
    & General & Reserved \\ 
    \cmidrule{2-3}
    Male & 74.91 & 12.08 \\ 
    Female & 11.74 & 1.27 \\ 
    \bottomrule
    \end{tabular}
\end{table}

In Table~\ref{tab:koo1} we show the \% breakup of the cross-sectional categories in the Koo dataset. We observe that the largest representation is from the general category males while the smallest is from the reserved category females. In the latest time point (see Table~\ref{tab:koo2}) we observe higher female representation than in the oldest time point (see Table \ref{tab:koo3}). The \% of females (both general and reserved) in top 1\% users sorted by followers is relatively larger than in the bottom 1\% followers (see Tables~\ref{tab:koo4} and \ref{tab:koo5}). This is exactly the opposite (see Tables~\ref{tab:koo4} and \ref{tab:koo5}) for males (both general and reserved). We believe that a possible reason could be that women have closed coteries of followership.

\subsection{Ethical implications}
Like any other classification task, it can also be potentially misused when in the hands of malicious actors. Instead of reduction of bias, the same technology can be used to enforce discrimination. Hence, we request the researchers to exercise caution while using this technology as some demography classification APIs are already publicly available. Further, to keep personally identifiable data private, we opensource the codebase to collect the datapoints instead of sharing the datasets, a policy ubiquitous for social science researchers.

\end{document}